\documentclass[twocolumn,showpacs,preprintnumbers,amsmath,amssymb]{revtex4}
\usepackage{graphicx}
\usepackage{bm}

\begin{document}

\title{Electron State Entanglement as Double Ionization Result }
\author{C.V. Usenko }
\email{usenko@ups.kiev.ua}
\author {V.O. Gnatovskyy}
\email{hnat@ups.kiev.ua}
\affiliation{Physics Department of Kyiv
Shevchenko University, Kyiv, Ukraine}
\begin{abstract}
A theoretical model allowing to explain the behavior of a correlation function
of electron pair in process of double-ionization has been built. It turnes out
the correlation functions' behavior can be explained by non-orthogonality
of entangled wave functions of electrons and the change of spin state of a
part of electron pairs in the process of atom double ionization from singlet
to triplet. It is shown that properties of correlation function are in the
best accordance with experimental data under assumption that the probability
of electrons' transition into triplet state is of the order of magnitude 0.5
ore higher.
\end{abstract}


\maketitle

\section{ Introduction}

Effect of double ionization has been described recently
\cite{b1,b2,b3,b4}. A characteristic feature of this phenomenon
is the simultaneous release of two electrons accompanied by
correlation of their momenta. The dependence of the correlation
function on the difference between the momenta of the electrons
is given in \cite{b1} for several series of experiments.

A significant error in measurements of the correlation function dependence on the difference between registered momenta allows to fit some theoretical models to the experimental data obtained. Some models of double ionization partially explaining the correlation of momenta of the released electrons were considered in papers [1,5], however the initial and final states of the electron pair were assumed to be singlet. In contrast to those reports we assume that some electron pairs undergo a singlet-triplet transition in the double ionization process.

A substantial correlation of the electron momenta found in the experiments indicates essential modification of the electron states, since, at least in helium, the electrons are in a singlet state and have identical momentum wave functions and therefore the calculated correlation of momenta is quite weak [1].
The entanglement of the momentum parts of electron wave functions in double ionization causing essential correlation of the monenta, can be explained, in our point of view, by a singlet-triplet transition in a part of electron pairs. Magnetic field created by the quickly moving charged particles (first of all, by nuclei), can change the spin state. Electrons in discrete states in the Coulomb field of a nucleus cannot make a transition to a triplet state, since it would lead to a substantial increase in energy (1s-2p transition), whereas there are no such restrictions for transitions between continuous states.

In this report we have shown that the non-orthogonality of entangled electron wave functions and the change of a spin state of electrons substantially influence the properties of the correlation function.
By neglecting the Coulomb repulsion of released electrons and, as a consequence, change of the correlation of momenta after ionization, we suppose that the electrons form an entangled pair either in spin (singlet state) or momentum (triplet state)
space and compare the calculated correlation function with the results of experiments.
Theory and experiment agree when singlet and triplet states are equally probable.

\section{Calculation of correlation function.}

Choose the wave functions of released electrons as
\begin{equation}
\varphi _{1,2}\left( \mathbf{p}_{1,2},\mathbf{P,}\widetilde{\mathbf{p}}%
,t\right)= \nonumber
\end{equation}
\begin{equation}
=\frac{1}{\left( 2\pi \sigma ^{2}\right) ^{3/4}}\exp \left( -\frac{%
\left( \mathbf{p}_{1,2}-\frac{\mathbf{P\pm }\widetilde{\mathbf{p}}}{2}%
\right) ^{2}}{4\sigma ^{2}}-i\frac{\mathbf{p}_{1,2}^{2}}{2m\hbar }t\right),
\label{e1}
\end{equation}
where $\mathbf{P}$ is the total average momentum of electrons, $\mathbf{\tilde{p%
}}$ their average momentum relative to the center of mass and $\sigma $ is an
uncertainty in momentum of each electron. Average momenta of electrons in
these states are
\begin{equation}
\langle \mathbf{p}_{1,2}\rangle =\frac{\mathbf{P\pm }\widetilde{\mathbf{p}}}{%
2}.
\end{equation}
Average coordinates of particles correspond to free motion
\begin{equation}
\langle \mathbf{r}_{1,2}\rangle =\frac{\left( \mathbf{P\pm }\widetilde{%
\mathbf{p}}\right) t}{2m}.
\end{equation}
An uncertainty in coordinate of each electron at the initial moment is equal to
$\frac{\hbar }{2\sigma }$ and increases if the electrons are assumed to be freely moving
\begin{equation}
\sigma _{\mathbf{r}}\left( t\right) =\frac{\hbar }{2\sigma }\sqrt{1+\frac{%
4\sigma ^{4}}{\hbar ^{2}m^{2}}t^{2}}.
\end{equation}

At the initial moment the electrons cannot be found at too large distances from the location of atom before ionization and, therefore, the uncertainty of coordinate is restricted. Thus, the uncertainty in momentum is restricted from below due to spatial localization of initial electron states.

Taking into account the change of spin state of electrons, the
spatial part of two-particle wave function in momentum
representation can be either symmetric or anti-symmetric
\begin{widetext}
\begin{equation}
\Psi _{S=0}\left( \mathbf{p}_{1},\mathbf{p}_{2};\mathbf{P,}
\widetilde{\mathbf{p}},t\right)= \frac{\varphi _{1}\left(
\mathbf{p}_{1};\mathbf{P,} \widetilde{\mathbf{p}},t\right)
\varphi _{2}\left( \mathbf{p}_{2};\mathbf{P,}
\widetilde{\mathbf{p}},t\right) +\varphi _{2}\left(
\mathbf{p}_{1};\mathbf{P, }\widetilde{\mathbf{p}},t\right)
\varphi _{1}\left( \mathbf{p}_{2};\mathbf{P,}
\widetilde{\mathbf{p}},t\right) }{\sqrt{2\left( 1+J^{2}\right) }},
\end{equation}


\begin{equation}
\Psi _{S=1}\left(
\mathbf{p}_{1},\mathbf{p}_{2};\mathbf{P,}\widetilde{
\mathbf{p}},t\right) =\frac{\varphi _{1}\left(
\mathbf{p}_{1};\mathbf{P,} \widetilde{\mathbf{p}},t\right)
\varphi _{2}\left( \mathbf{p}_{2};\mathbf{P,}
\widetilde{\mathbf{p}},t\right) -\varphi _{2}\left(
\mathbf{p}_{1};\mathbf{P, }\widetilde{\mathbf{p}},t\right)
\varphi _{1}\left( \mathbf{p}_{2};\mathbf{P,
}\widetilde{\mathbf{p}},t\right) }{\sqrt{2\left( 1-J^{2}\right)
}}.
\end{equation}
\end{widetext}

 These functions are normalized and satisfy the
Schrodinger non-stationary equation
\begin{equation}
i\hbar \frac{{\partial \Psi _{S}}}{{\partial t}}=H\Psi _{S} , \label{e11}
\end{equation}
where Hamiltonian describes a pair of free particles $H=\frac{\mathbf{p}%
_{1}^{2}+\mathbf{p}_{2}^{2}}{2m}$.

An overlap integral for $\varphi _{1}$and $\varphi _{2}$ is
\begin{eqnarray}
J=\int \varphi _{1}^{\ast }\left(
\mathbf{p;P,}\widetilde{\mathbf{p}},t\right) \varphi _{2}\left(
\mathbf{p;P,}\widetilde{\mathbf{p}},t\right) d \mathbf{p}=
\nonumber
\\
=\exp \left( -\frac{\widetilde{p}^{2}}{8\sigma ^{2}}\right)
\label{e33}
\end{eqnarray}
and depends on average momentum of particles relative to the
center of mass only.

Let us consider the statistical characteristics of double ionization
process. We assume that after ionization the electrons gain a total average
momentum $\mathbf{P}$ , the difference between average momenta being $\mathbf{%
\tilde{p}}$ and total spin of system $S$ . We denote the differential
cross-section of this process as

\begin{equation}
\Theta _{S}\left( \mathbf{P,}\widetilde{\mathbf{p}}\right) =\frac{d\sigma
_{S}\left( \mathbf{P,}\widetilde{\mathbf{p}}\right) }{d\mathbf{P}d\widetilde{%
\mathbf{p}}}.
\end{equation}

Average characteristics of the state are $\mathbf{P}$ , $\mathbf{\tilde{p}}$
and $S$ . Their estimate can be made taking into account the statistical results of measurements.

Then the electrons freely move towards detectors and are registered with some
momenta $\mathbf{p}_{1}$ and $\mathbf{p}_{2}$ that are not equal to average
electron momenta $\frac{\mathbf{P\pm }\widetilde{\mathbf{p}}}{2}$ and can be
very different from the average values. For each specific state of an electron
pair the differential ionization cross-section $\Phi _{S}\left( {\mathbf{p}%
_{1},\mathbf{p}_{2};\mathbf{P},\mathbf{\tilde{p}},t}\right) $ with
registration of fixed momenta $\mathbf{p}_{1}$ and $\mathbf{p}_{2}$ through
the channel with average electrons' momenta $\langle \mathbf{p}_{1,2}\rangle
=\frac{\mathbf{P\pm }\widetilde{\mathbf{p}}}{2}$\ and total spin $S$ is
equal to the product of the differential cross-section of ionization into this
state and probability of registering momenta $\mathbf{p}_{1}$ and $\mathbf{p}%
_{2}$ in this state. This probability is determined by the square of
two-particle wave function modulus, so the differential ionization cross-section
through the channel with average momenta $\frac{\mathbf{P\pm }\widetilde{%
\mathbf{p}}}{2}$, total spin $S$ and registered momenta $\mathbf{p}_{1}$ and
$\mathbf{p}_{2}$ is
\begin{eqnarray}
\Phi _{S}\left( {\mathbf{p}_{1},\mathbf{p}_{2};\mathbf{P},\mathbf{\tilde{p}}%
,t}\right) =\nonumber
\\
=\Theta _{S}\left( {\mathbf{P},\mathbf{\tilde{p}}}\right) \left| {%
\Psi _{S}\left( {\mathbf{p}_{1},\mathbf{p}_{2};\mathbf{P},\mathbf{\tilde{p}}%
,t}\right) }\right| ^{2}.  \label{e10}
\end{eqnarray}

A differential ionization cross-section with registering fixed momenta $%
\mathbf{p}_{1}$ and $\mathbf{p}_{2}$ is determined by the sum over all
possible states and equals to the sum of integrals of products of $\Theta
_{S}\left( {\mathbf{P},\mathbf{\tilde{p}}}\right) $ and the squared modulus of
a two-particle wave function $\left| {\Psi _{S}\left( {\mathbf{p}_{1},\mathbf{p%
}_{2};\mathbf{P},\mathbf{\tilde{p}},t}\right) }\right| ^{2}$, this sum is
taken over spin states:
\begin{eqnarray}
\Phi \left( {\mathbf{p}_{1},\mathbf{p}_{2}}\right)=\sum\limits_{S}{\frac{{%
d\sigma _{S}\left( {\mathbf{p}_{1},\mathbf{p}_{2}}\right) }}{{d\mathbf{p}%
_{1}d\mathbf{p}_{2}}}} =\nonumber
\\
=\sum\limits_{S}{\int {\Theta _{S}\left( {\mathbf{P},%
\mathbf{\tilde{p}}}\right) }}{\left| {\Psi _{S}\left( {\mathbf{p}_{1},\mathbf{p%
}_{2};\mathbf{P},\mathbf{\tilde{p}},t}\right) }\right| }^{2}d\mathbf{P}d%
\mathbf{\tilde{p}}.  \label{e3}
\end{eqnarray}

Now we can determine the cross-section of ionization with registration of a
momentum $\mathbf{p}$ for one of the two electrons and the total cross-section for double ionization:
\begin{equation}
\rho \left( \mathbf{p}\right)  =\int {\Phi \left( {\mathbf{p},\mathbf{
p^{\prime }}}\right) d\mathbf{p^{\prime }}}, \label{e12}
\end{equation}
\begin{eqnarray}
N_{tot} =\int {\Phi \left( {\mathbf{p},\mathbf{p^{\prime
}}}\right) d \mathbf{p}d\mathbf{p^{\prime }}}=\nonumber
\\
= \sum\limits_{S}{\int {\Theta _{S}\left( {
\mathbf{P},\mathbf{\tilde{p}}}\right) }}\left| {\Psi _{S}\left(
{\mathbf{p}, \mathbf{p^{\prime
}};\mathbf{P},\mathbf{\tilde{p}}}\right) }\right| ^{2}d
\mathbf{P}d\mathbf{\tilde{p}}d\mathbf{p}d\mathbf{p^{\prime
}}=\nonumber
\\
=\sum\limits_{S}{\int {\Theta _{S}\left( {\mathbf{P},\mathbf{\tilde{p}}}%
\right) d\mathbf{P}d\mathbf{\tilde{p}}}}=N_{0}+N_{1},  \label{e13}
\end{eqnarray}
where $N_{0}=\int {\Theta
_{0}\left( {\mathbf{P},\mathbf{\tilde{p}}}\right) d\mathbf{P}d\mathbf{\tilde{%
p}}}$ and $N_{1}=\int {\Theta _{1}\left( {\mathbf{P},\mathbf{\tilde{p}}}%
\right) d\mathbf{P}d\mathbf{\tilde{p}}}$ are the total ionization
cross-sections into singlet and triplet states respectively. This gives us a possibility of obtaining measurable characteristics – intensity of registration of two electrons emitted in the same act of double ionization
\begin{equation}
I_{cor}\left( {\Delta p}\right) =\left( {\Delta p}\right) ^{2}\int {\Phi
\left( {\mathbf{p}_{1},\mathbf{p}_{1}+\Delta \mathbf{p}}\right) d\mathbf{p}%
_{1}d\mathbf{n}}  \label{e14}
\end{equation}
along with intensity of registration of two electrons irrespective of the fact whether they were emitted in the same or different acts of double ionization
\begin{equation}
I_{uncor}\left( {\Delta p}\right) =\frac{{\left( {\Delta p}\right) ^{2}}}{N_{tot}}%
\int {\rho \left( {\mathbf{p}_{1}}\right) \rho \left( {\mathbf{p}_{1}+\Delta
\mathbf{p}}\right) d\mathbf{p}_{1}d\mathbf{n,}}  \label{e15}
\end{equation}
that determine the correlation function

\begin{equation}
R\left( {\Delta p}\right) =\frac{{I_{cor}\left( {\Delta p}\right) }}{{%
I_{uncor}\left( {\Delta p}\right) }}-1,  \label{e16}
\end{equation}
where $\Delta p=\left| {\mathbf{p}_{1}-\mathbf{p}_{2}}\right| $ is the
difference of momenta of registered electrons. It is this function that is
observed in experiments on double ionization.

The differential ionization cross-section $\Phi \left( {\mathbf{%
p}_{1},\mathbf{p}_{2}}\right) $ and the intensities $I_{cor}$ and $I_{uncor}$
depend not only on two-particle wave function, but on the probability of
ionization through a concrete state of the electron pair as well, and this fact is
represented by the factor $\Theta _{S}\left( {\mathbf{P},\mathbf{\tilde{p}}}%
\right) $ in (\ref{e3},\ref{e12},\ref{e14},\ref{e15}). If the only bossible state of electrons with certain
average momenta and fixed spin is formed in the process of ionization then the
integration over states disappears. Therefore this factor is replaced by the
number of registered events of double ionization $N_{tot}$  and the correlation
function is determined by a two-particle wave function in this state. If a pair of electrons with the same probabilities $%
P_{S}\left( {\mathbf{P},\mathbf{\tilde{p}}}\right) $ undergoes a transition to a state with total spin $S$ and average momenta $\left\langle {\mathbf{p}_{1,2}}%
\right\rangle =\frac{{{{\mathbf{P}\pm \mathbf{\tilde{p}}}}}}{2}$ in the process of ionization, the
differential ionization cross-section though each of those states is equal
to $\Theta _{S}\left( {\mathbf{P},\mathbf{\tilde{p}}}\right)
=N_{tot}P_{S}\left( {\mathbf{P},\mathbf{\tilde{p}}}\right) $. As a consequence, the correlation function does not depend only on the properties of a wave function of single state of an electron pair - it is a characteristic of ionization process average over all possible states.

The most essential experimental data can be summarized as follows. Firstly, alternation of sings (negative for a small difference of momenta, positive for momenta comparable with the atomic unit, and negative for large differences of momenta); secondly, a maximum lies between 2 and 3 a.u.

In the general expression for differential ionization cross-section obtained
above there are no assumptions about character of electron motion from a point
of ionization to detectors. If we assume that the electrons move freely, the following simplifications can be introduced.
First of all, in the momentum representation the square of modulus of a two-particle wave function is time-independent and the correlation function does not depend on the time interval necessary for the electrons to cover the distance from a point of ionization to detectors. With the aim of further
simplifications in the expression (\ref{e3}) let's consider the fact that
according to (\ref{e1}) the squared modulus of a two-particle wave function
is almost independent on average momenta $\mathbf{P}$ and $\mathbf{\tilde{%
p}}$ as long as their values are smaller then momentum uncertainty of each
electron $\sigma $ . By assuming that after ionization the electrons
have such an average momenta we can replace the differential ionization
cross-section with registering electrons' momenta $\mathbf{p}_{1}$ and $%
\mathbf{p}_{2}$ by expression
\begin{equation}
\Phi \left( {\mathbf{p}_{1},\mathbf{p}_{2}}\right) =\sum\limits_{S}N_{S}{%
\left| \Psi _{S}{\left( {\mathbf{p}_{1},\mathbf{p}_{2},t}\right) }\right|
^{2},}  \label{e17}
\end{equation}
where integration over momenta of resulting states gives only the total
ionization cross-sections with spins $S=0$ and $S=1$ . Let's denote the
probability of a singlet-triplet transition as $f$ .
In accordance with (\ref{e13}) the differential ionization cross-sections can be written as
 $N_{0}=\left( 1-f\right) N_{tot}$ , $N_{1}=fN_{tot}$ .
A differential ionization cross-section with registering of fixed momenta of
electrons
\begin{eqnarray}
\Phi \left( {\mathbf{p}_{1},\mathbf{p}_{2}}\right) =\nonumber
\\
=N_{tot}\left( {f\left| {%
\Psi _{1}\left( {\mathbf{p}_{1},\mathbf{p}_{2},t}\right) }\right|
^{2}+\left( {1-f}\right) \left| {\Psi _{0}\left( {\mathbf{p}_{1},\mathbf{p}%
_{2},t}\right) }\right| ^{2}}\right) ,  \label{e18}
\end{eqnarray}
is a product of the total cross-section and sum of probabilities
of registering electrons with fixed momenta in singlet and
triplet states. Thus, if we assume that average momenta of
electrons are small in comparison with uncertainties of their
momenta, the probability of ionization through a triplet state
will be determined by the product of a triplet wave function and
probability of spin flip, $f$. Then the uncertainty of momentum
of electrons $\sigma $ acts as a scaling factor, and the shape of
correlation function is mainly determined by the ratio of
probabilities of ionization producing entangled pairs in singlet
and triplet states.

\begin{figure}[h]
\includegraphics[scale=0.7]{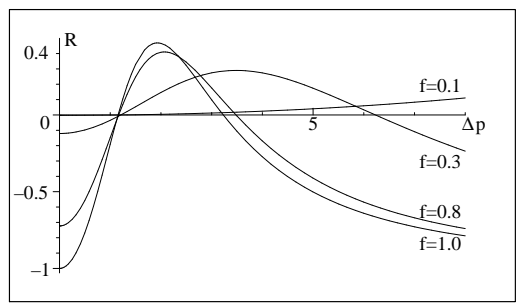}
\caption{\label{f1}Correlation functions for processes of
ionization into states with different probabilities of transition
to triplet state.} \label{fg1}
\end{figure}

A correlation function for a process of ionization into states
with the same values of momentum uncertainty and different
probabilities of transition to a triplet state is shown in Fig.
\ref{fg1}. These graphs show that the correlation of momenta
decreases along with the
 probability of transition into a triplet state. It is clear that the spatial part of a two-particle wave function will be a product of one-particle wave functions if average momenta of electrons are identical.
 As a consequence, the squared
modulus of a two-particle wave function is
equal to $\left| {\Psi _{0}\left( {\mathbf{p}_{1},\mathbf{p}_{2},t}\right) }%
\right| ^{2}=\left| {\varphi \left( {\mathbf{p}_{1},t}\right) }\right|
^{2}\left| {\varphi \left( {\mathbf{p}_{2},t}\right) }\right| ^{2}$
corresponding to an independent distribution of two random values.

Now we consider the dependence of the correlation function on the average momenta of electrons. We assume that the value of  $\Theta _{S}$ differs sufficiently from
zero in some area near average momenta $\mathbf{P}$, $\mathbf{\tilde{p}}$,
 and the width of this area is noticeably smaller than
 the uncertainty in momentum $\sigma$. In such an area a two-particle wave function
$\Psi _{S}$ is essentially constant and the differential cross-section of ionization with registering of momenta $\mathbf{p}_{1}$ and $\mathbf{p}_{2}$ of electrons can approximately be replaced by
\begin{equation}
\Phi \left( {\mathbf{p}_{1},\mathbf{p}_{2}}\right) =\sum\limits_{S}{%
N_{S}\left| {\Psi _{S}\left( {\mathbf{p}_{1},\mathbf{p}_{2};\mathbf{P},%
\mathbf{\tilde{p}},t}\right) }\right| ^{2}},
\end{equation}
where $N_{S}=\int {\Theta _{S}\left( {\mathbf{P},\mathbf{\tilde{p}}}\right) d%
\mathbf{P}d}\mathbf{\tilde{p}}$ again stands for the ionization
cross-sections through the channel with spin $S$ .

Under these assumptions the properties of the correlation function $R\left( {%
\Delta p,\mathbf{\tilde{p}}}\right) $ depend on the difference of average
momenta of electrons $\mathbf{\tilde{p}}$ and do not depend on the total
average momentum $\mathbf{P}$.

The intensities will depend on $\mathbf{\tilde{%
p}}$ as

\begin{widetext}
\begin{equation}
I_{cor}=N_{tot}\left( \Delta p\right) ^{2}\frac{\exp \left( -\frac{\left(%
\Delta p\right) ^{2}}{4\sigma ^{2}}\right) J^{2}}{2\sqrt{\pi }\sigma ^{3}} %
\left( \left( 1-f\right) \frac{\frac{\sinh z}{z}+1}{1+J^{2}}+f\frac{\frac{%
\sinh z}{z}-1}{1-J^{2}}\right)   \label{e19}
\end{equation}

\begin{equation}
I_{uncor}=N_{tot}\left( \Delta p\right) ^{2}\frac{\exp \left( -\frac{%
\left( \Delta p\right) ^{2}}{4\sigma ^{2}}\right)
J^{2}}{4\sqrt{\pi }\sigma ^{3}}\left( A+B+C\right) ,  \label{e20}
\end{equation}
\end{widetext}
\begin{eqnarray}
A =\left( 1-f\right) ^{2}\frac{1+2J^{4}+J^{2}\frac{\sinh z}{z}+8J^{5/2}%
\frac{\sinh z/2}{z}}{\left( 1+J^{2}\right) ^{2}},  \nonumber \\
B =f^{2}\frac{1+2J^{4}+J^{2}\frac{\sinh z}{z}-8J^{5/2}\frac{\sinh z/2}{z}}{%
\left( 1-J^{2}\right) ^{2}},  \nonumber \\
C =2f\left( 1-f\right) \frac{1-2J^{4}+J^{2}\frac{\sinh
z}{z}}{\left( 1+J^{2}\right) \left( 1-J^{2}\right) }.  \nonumber
\end{eqnarray}

Now we consider the dependence of correlation function on $\mathbf{\tilde{%
p}}$
 separately for a singet and triplet state. The
correlation function for each of the states depends on expression $z=\frac{%
\widetilde{p}\Delta p}{2\sigma ^{2}}$ and the overlap integral (\ref{e33})

\begin{equation}
R_{0}\left( \Delta p,\widetilde{p}\right) =\frac{2\left( \frac{\sinh z}{z}%
+1\right) \left( 1+J^{2}\right) }{1+2J^{4}+J^{2}\frac{\sinh z}{z}+8J^{5/2}%
\frac{\sinh z/2}{z}}-1,  \label{e21}
\end{equation}

\begin{equation}
R_{1}\left( \Delta p,\widetilde{p}\right) =\frac{2\left( \frac{\sinh z}{z}%
-1\right) \left( 1-J^{2}\right) }{1+2J^{4}+J^{2}\frac{\sinh z}{z}-8J^{5/2}%
\frac{\sinh z/2}{z}} -1.  \label{e22}
\end{equation}

\begin{figure}[h]
\includegraphics[scale=0.7]{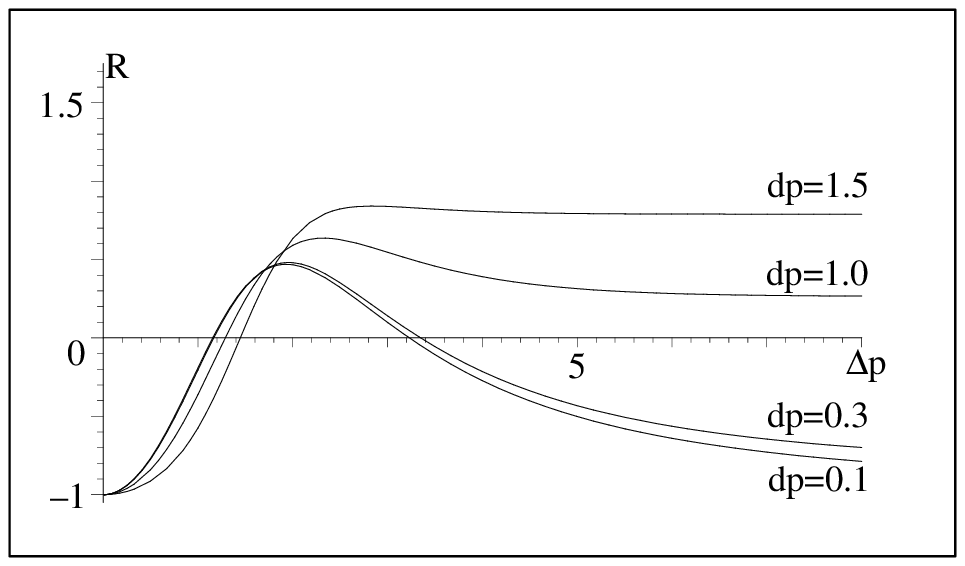}\\
\caption{Correlation functions for several values of difference
of momenta for electrons in triplet state.} \label{f2a}
\end{figure}

\begin{figure}[h]
\includegraphics[scale=0.7]{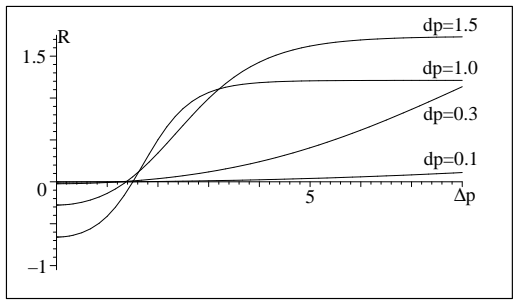}
\caption{Correlation functions for several values of difference
of momenta for electrons in singlet state.}\label{f2b}
\end{figure}

Graphs of correlation function for several values of $\mathbf{\tilde{%
p}}$ in a triplet state are given in Fig. \ref{f2a}.
 They show that an
increase in $\mathbf{\tilde{p}}$ affects only the
character of correlation but the state itself remains strongly
correlated. Starting from $\tilde{p}\simeq \sigma $ the correlation
function loses its second null and remains positive for large $\Delta p$,
which contradicts the experimental data [1].

Dependence of correlation function on $\mathbf{\tilde{p}}$ for a
singlet state is shown in Fig. \ref{f2b}. In this state, on the
contrary, the correlation increases along with the difference of
average momenta as the influence of symmetrization of
one-particle wave functions becomes noticeable. Nevertheless, the
correlation function has only a single null.

These graphs indicate that properties of correlation function of
singlet and triplet states are different. Such distinctions are most visible for large differences in the momenta registered.
When the difference
of average monenta of electrons is small $\left( \widetilde{p}\ll \sigma
\right) $, the correlation function is negative only for a triplet state.
 In the area of small differences of momenta registered $\Delta p\simeq 1a.u.$
 the correlation functions of a singlet and triplet
state are negative. For a triplet sate the correlation function is
equal to -1 at zero, whereas in singlet state it depends on $\widetilde{p}$\ and can tend to zero.

\section{Comparison to experiment.}
\begin{figure}[h]
\begin{center}
\includegraphics[scale=0.7]{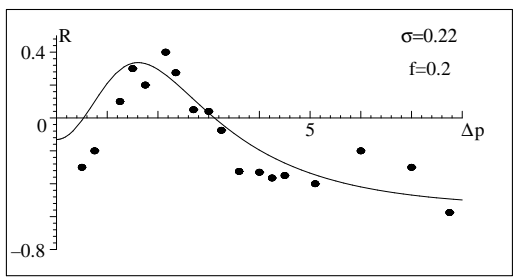}
\end{center}
\caption{Approximation of experimental data for ionization of He by $Au^{53+}$.}
\label{f3a}
\end{figure}

\begin{figure}[h]
\begin{center}
\includegraphics[scale=0.7]{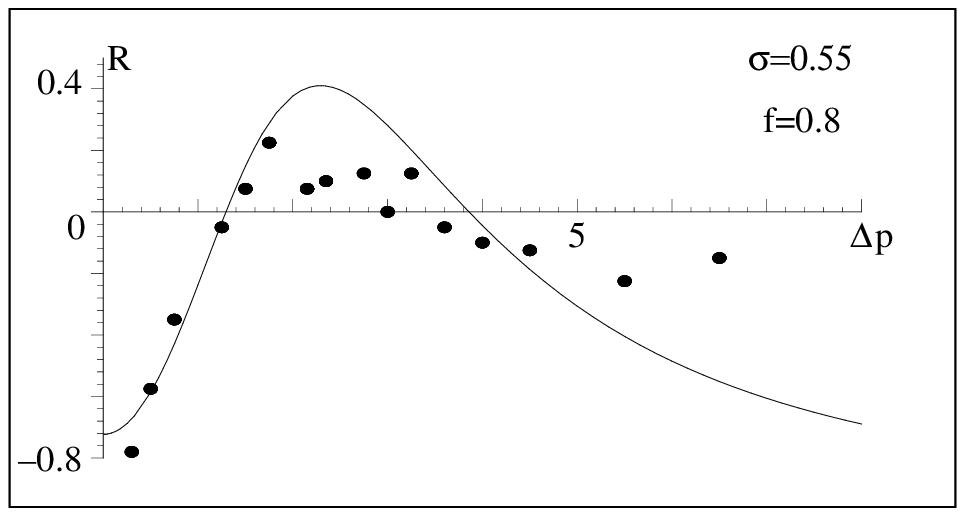}
\end{center}
\caption{Approximation of experimental data for ionization of He by $C^{6+}$}
\label{f3b}
\end{figure}

\begin{figure}[h]
\begin{center}
\includegraphics[scale=0.7]{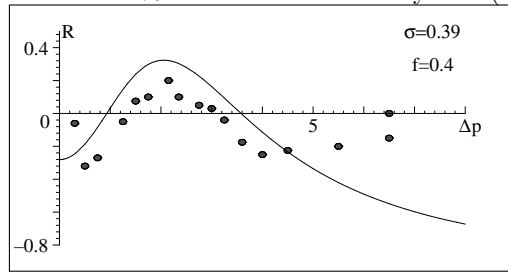}
\end{center}
\caption{Approximation of experimental data for ionization of Ne by$Au^{53+}$.}
\label{f3c}
\end{figure}

The uncertainty of electrons' momenta
produced after double ionization should depend on the process of
ionization. This uncertainty can be determined from experimental results.

The least square approximation of the experimental data given in
[1] leads to an uncertainty of 0.22 a.u. and error of
approximation of 15\% for ionization of He by Au$^{53+}$ (Fig.
\ref{f3a}), uncertainty of 0.55 a.u. and error of approximation
of 19\% for ionization of He by C$^{6+}$ (Fig. \ref{f3b}), and
uncertainty of 0.39 a.u. and error of approximation of 18\% for
ionization of Ne by Au$^{+53}$ (Fig. \ref{f3c}).

\section{Conclusions.}

We have considered  the influence of a spin state change of an electron pair in process of double ionization on the correlation function properties.
Taking into account a large scatter of experimental data-points, the results of our approximation show that experiments on double ionization are in a good agreement with the assumption about the change of the spin state of a part of the electron pairs. The negative sign of the correlation function for large momenta differences together with the significant correlation for small momenta differences support our assumption. The effect of spin flip in double ionization needs further experimental investigations.

\end{document}